\documentclass[aip,jcp,amsmath,amssymb,reprint,numerical]{revtex4-1}

\usepackage{graphicx}% Include figure files
\usepackage{dcolumn}% Align table columns on decimal point
\usepackage{bm}% bold math
\usepackage[caption=false]{subfig}

\begin{document}

\title{Production of carbon clusters $\text{C}_3$ to $\text{C}_{12}$ with a cryogenic buffer-gas beam source}

\author{C.~J.~E.~Straatsma}
\email{cameron.straatsma@colorado.edu}
\affiliation{JILA and Department of Electrical, Computer, and Energy Engineering, University of Colorado, Boulder, Colorado 80309-0440}

\author{M.~I.~Fabrikant}
\affiliation{JILA and Department of Physics, University of Colorado, Boulder, Colorado 80309-0440}

\author{G.~E.~Douberly}
\affiliation{Department of Chemistry, University of Georgia, Athens, Georgia 30602-2556}

\author{H.~J.~Lewandowski}
\affiliation{JILA and Department of Physics, University of Colorado, Boulder, Colorado 80309-0440}

\date{\today}

\begin{abstract}
Cryogenic buffer-gas beam sources are capable of producing intense beams of a wide variety of molecules, and have a number of advantages over traditional supersonic expansion sources. In this work, we report on a neon matrix isolation study of carbon clusters produced with a cryogenic buffer-gas beam source. Carbon clusters created by laser ablation of graphite are trapped in a neon matrix and detected with a Fourier transform infrared spectrometer in the spectral range $4000-1000~\text{cm}^{-1}$. Through a study of carbon cluster production as a function of various system parameters, we characterize the behavior of the buffer-gas beam source and find that approximately $10^{11}-10^{12}$ of each cluster is produced with each pulse of the ablation laser. These measurements demonstrate the usefulness of cryogenic buffer-gas beam sources for producing molecular beams of clusters.
\end{abstract}

\maketitle

\section{\label{sec:introduction}Introduction}
Laser vaporization based cluster sources are widely used in molecular spectroscopy.~\cite{Duncan2012} For example, they have been used to study a variety of metal atom clusters,~\cite{Dietz1981,Bondybey1981,DeHeer1993} metal alloy clusters,~\cite{Rohlfing1984a,Wheeler1988} and clusters of metal oxides~\cite{Deng1996,Foltin1999} and carbides.~\cite{Guo1992,Pilgrim1993} Additionally, the laser vaporization cluster source has found widespread use in the study of linear chains and rings of carbon atoms,~\cite{Rohlfing1984b,Heath1987,Belau2007} including the discovery of the $\text{C}_{60}$ buckminsterfullerene.~\cite{Kroto1985} The majority of these laser vaporization sources consist of a solid target held in the vicinity of a pulsed valve from which a gas pulse is introduced to cool ablated molecules via entrainment in a supersonic expansion. A growth channel located downstream of the pulsed valve is often used to increase collisional interaction time and cluster yield. Given the wide applicability of these sources, it is of general interest to investigate other methods for producing clusters, such as the cryogenic buffer-gas beam source,~\cite{Maxwell2005,Patterson2009,Lu2011,Hutzler2012} which could provide a cluster beam with increased intensity and lower internal state temperature compared to a traditional supersonic expansion source.

Both supersonic expansion and buffer-gas beam sources rely on collisions with a dense, cold buffer gas to cool hot molecules from $>1000~\text{K}$ into the few kelvin regime; however, their construction is significantly different. In its simplest form, a buffer-gas beam source consists of a cell with a constant flow of inert buffer gas (e.g., helium or neon) entering through a small aperture. The cell is held at a temperature on the order of $10~\text{K}$ to cool the buffer gas, into which the molecules of interest are introduced through laser ablation of a solid target located inside the cell. The ablated molecules become entrained in the flow of cold buffer gas where they rapidly cool and eventually exit the cell through a small aperture to form a bright beam of cold molecules.

Buffer-gas beam sources have a number of advantages over traditional supersonic expansion sources, such as lower forward velocity, higher brightness, and a longer collisional cooling timescale.~\cite{Hutzler2012} Both sources readily achieve peak buffer gas densities on the order of $10^{17}~\text{cm}^{-3}$; however, molecules ablated in the vicinity of a supersonic expansion typically interact with buffer gas of this density for times on the order of $100~\mu\text{s}$,~\cite{Duncan2012} whereas this timescale can be longer than $1~\text{ms}$ inside a buffer-gas cell.~\cite{Hutzler2012} This longer interaction time with a high density, cold buffer gas could lead to lower internal-state temperatures for many molecules compared to a supersonic expansion source. An additional benefit of the longer interaction time in a buffer-gas source is the possibility of lower internal state temperatures for large, strongly bound clusters, such as carbon clusters or clusters of metal oxides and carbides, all of which experience significant heating upon aggregation due to binding energies of $4-5~\text{eV}$ per bond.~\cite{Duncan2012} To date, many molecular species have been successfully cooled with a buffer-gas beam source, including diatomic~\cite{Maxwell2005,Hutzler2011,Barry2011,Skoff2011,Bu2017,Iwata2017} and polyatomic~\cite{VanBuuren2009,Sommer2009,Li2016,Spaun2016,Tokunaga2017} molecules, but there currently exist no studies on the formation and cooling of cluster systems with such a source. The ability to cool a wide range of clusters to their vibrational ground state could aid in their structural determination with high-resolution spectroscopy.

In this work, we investigate the production of carbon clusters with a neon buffer-gas beam source, which is in contrast to previous studies of small carbon clusters~\cite{VanOrden1998} that primarily utilized supersonic expansion sources. Using matrix isolation spectroscopy, we observe clusters ranging in size from $\text{C}_3$ to $\text{C}_{12}$, and estimate the number of each cluster produced. Clusters in this size range have been previously produced using a supersonic expansion source with a small growth channel.~\cite{Heath1991,Neubauer-Guenther2003,Belau2007} This paper is organized as follows. Section~\ref{sec:experimental} provides an overview of the apparatus and the experimental procedure used to characterize the buffer-gas beam source. Section~\ref{sec:results} investigates the production of carbon clusters as a function of various system parameters, and describes a method for quantifying the number of each cluster produced. Section~\ref{sec:conclusion} concludes and provides a future outlook.

\section{\label{sec:experimental}Experimental Methods}
\begin{figure}
\centering
\includegraphics[scale=1]{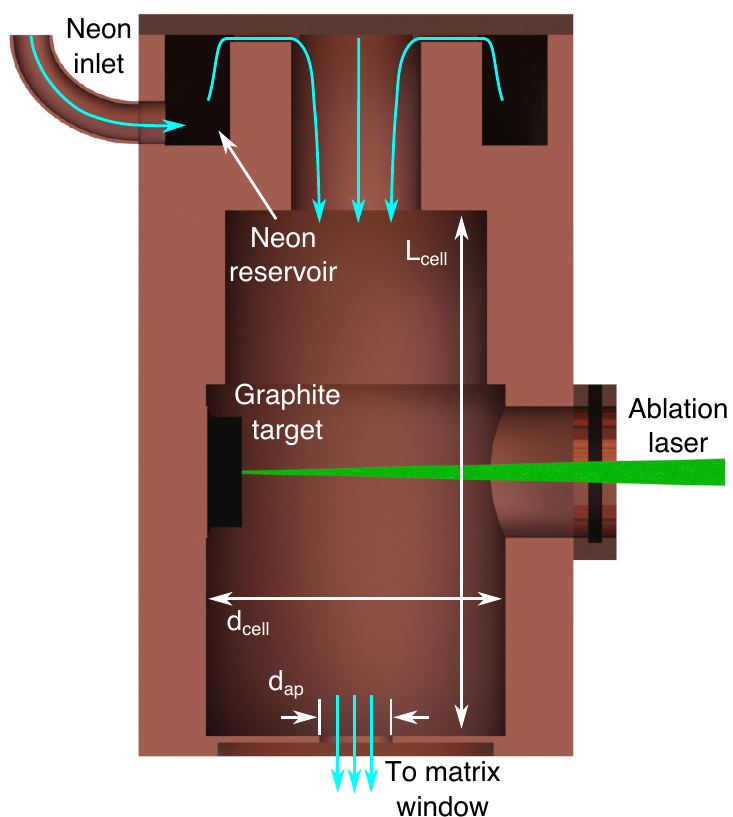}
\caption{\label{fig:apparatus} A schematic of the buffer-gas cell used for the production of carbon clusters. It consists of a copper box with a cylindrical bore ($L_{\text{cell}} = 40~\text{mm}$ and $d_{\text{cell}} = 20~\text{mm}$) drilled through the center. A thin plate attached to the bottom contains an aperture of diameter $d_{\text{ap}} = 5.3~\text{mm}$. Neon gas (blue arrows) first enters a reservoir at the top of the cell before flowing into the main cell, and exits through the aperture at the bottom. The ablation laser enters through an uncoated N-BK7 window attached to the outside of the cell, and is focused onto a graphite target attached to the inside of the cell.}
\end{figure}

A schematic of the buffer-gas cell used in this work is shown in Fig.~\ref{fig:apparatus}. 
It consists of a copper box with a cylindrical bore of length $L_{\text{cell}} = 40~\text{mm}$ and diameter $d_{\text{cell}} = 20~\text{mm}$. The flow of neon buffer gas into the cell is controlled by a mass flow controller (Alicat Scientific, MC-100SCCM-D), and first enters a small reservoir at the top of the cell to promote laminar flow through the main compartment of the cell. A thin aperture of diameter $d_{\text{ap}} = 5.3~\text{mm}$ located in the bottom face of the cell allows for beam extraction. Through a weak connection to the second stage of a two-stage cryocooler (SHI Cryogenics Group, RDK-415D), the cell is held at a temperature of $T_{\text{cell}} = 25~\text{K}$ to prevent neon gas from freezing to the cell walls. Note that the neon gas line is also anchored to the first stage of the cryocooler to cool the neon gas from room temperature to approximately $30~\text{K}$ prior to it entering the buffer-gas cell.

Carbon clusters are produced inside the buffer-gas cell by laser ablation of a graphite target located at the edge of the cylindrical bore halfway along the length of the cell. To ablate graphite, the output of a Nd:YAG laser (Spectra Physics, Quanta-Ray Lab-150, $532~\text{nm}$) is focused onto the target to a spot roughly $30~\mu\text{m}$ in diameter using a biconvex lens with a $400~\text{mm}$ focal length. To prevent the ablation laser from drilling a hole into the target,~\cite{Duncan2012} the beam is constantly scanned in two dimensions using a motorized mirror mount. Without scanning, we observe rapid degradation in the number of molecules produced per laser pulse.

Following ablation, the carbon clusters undergo collisions with the cold neon buffer gas for a few milliseconds before becoming entrained in the molecular beam extracted from the cell. The properties of this buffer-gas beam are determined by the cell geometry, the aperture diameter, and the flow rate of neon gas into the cell.~\cite{Hutzler2012} The first two variables are fixed in this experiment, whereas the neon flow rate can be varied in the range $0-100~\text{sccm}$ using the mass flow controller. Notably, both cluster formation and collisional cooling of the ablation products occur during the interaction time inside the cell.

\begin{figure*}
\centering
\subfloat{\includegraphics[scale=1]{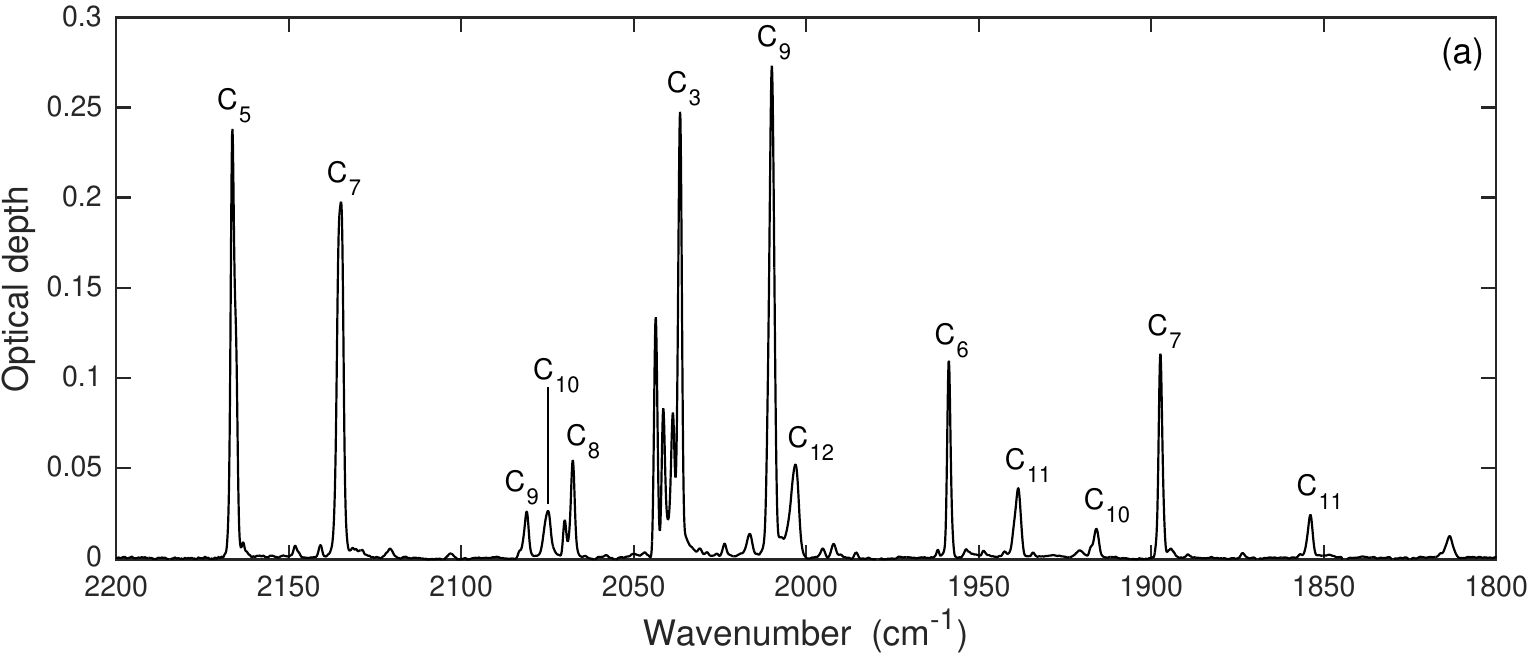}}\\
\subfloat{\includegraphics[scale=1]{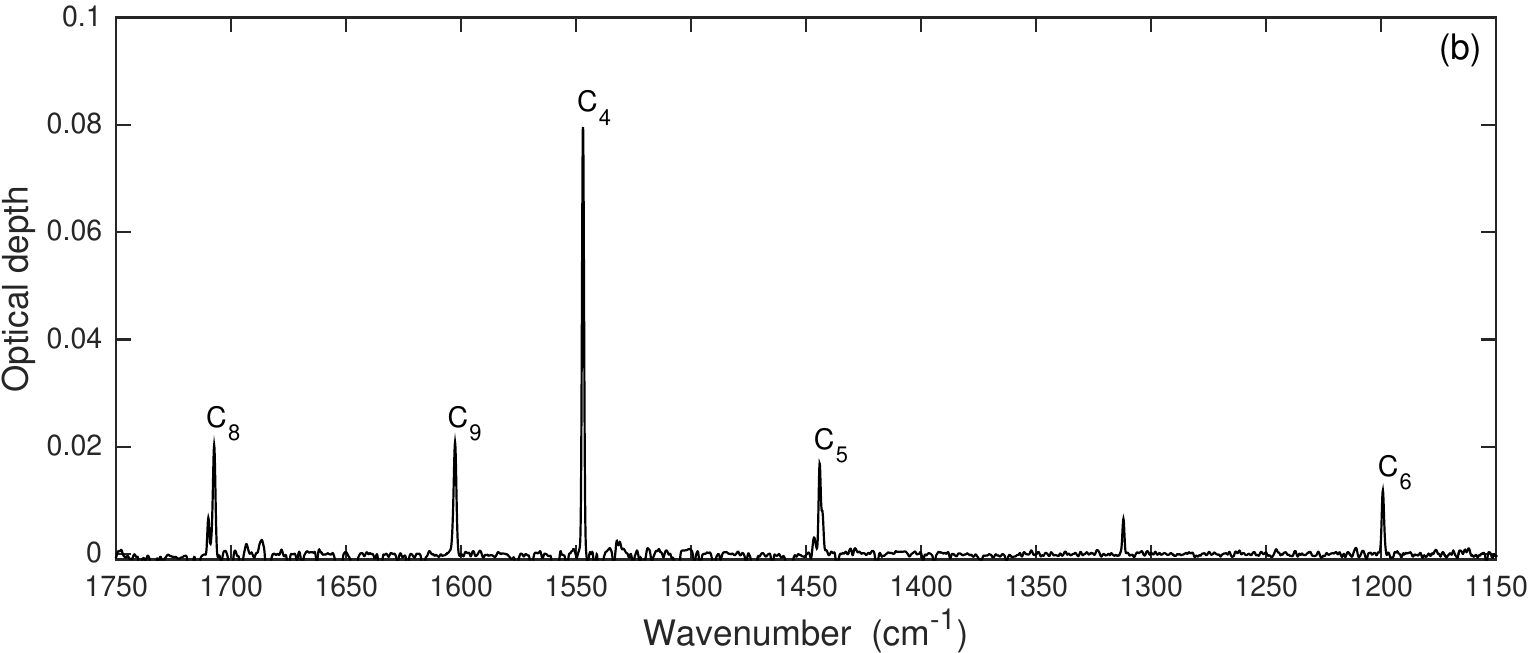}}
\caption{\label{fig:spectrum} Infrared absorption spectrum of carbon clusters isolated in a neon matrix at 5 K. The system parameters were $Q_{\text{Ne}} = 40~\text{sccm}$, $E_p = 2~\text{mJ}$, and $f_r = 7.5~\text{Hz}$. Note the difference in the vertical scale for the spectral range (a) $2200 - 1800~\text{cm}^{-1}$ and (b) $1750 - 1150~\text{cm}^{-1}$.}
\end{figure*}

To measure the number and composition of carbon clusters produced inside the neon buffer-gas cell, we use a matrix isolation spectroscopy setup.~\cite{Dunkin1998} The buffer-gas beam exiting the cell is frozen onto a $\text{CaF}_2$ window ($T\sim5~\text{K}$) located approximately $2~\text{cm}$ from the cell aperture. Infrared absorption spectra of matrix isolated carbon clusters are acquired in the range $4000-1000~\text{cm}^{-1}$ at $1~\text{cm}^{-1}$ resolution using a Fourier-transform infrared spectrometer (Thermo Fisher Scientific, Nicolet iS50). We average 36 scans for each acquired spectrum resulting in a dynamic range of approximately $70~\text{dB}$. This leads to a signal-to-noise ratio of about $925:1$ for the $\nu_6$ mode of $\text{C}_9$ as depicted in the spectrum of Fig.~\ref{fig:spectrum}(a), which is acquired after 30 minutes of deposition.

For a given set of system parameters, which include the neon flow rate $Q_{\text{Ne}}$, laser pulse energy $E_p$, and laser repetition rate $f_{r}$, we ablate continuously for two minutes then acquire a spectrum. During the acquisition of each spectrum, the neon flow is turned off and the ablation laser is physically blocked by a mechanical shutter. This process is repeated 15 times for a total deposition time of 30 minutes to determine the rate of increase in optical depth (i.e., absorbance) for each observed peak. Given this change in optical depth, we quantify the performance of our buffer-gas beam source by estimating the number of each carbon cluster produced per pulse of the ablation laser. The following section discusses the results of this experiment as a function of $Q_{\text{Ne}}$, $E_p$, and $f_{r}$, and describes our method for estimating molecule production per pulse from the rate of increase in optical depth.

\section{\label{sec:results}Results and Discussion}
A typical infrared spectrum of carbon clusters isolated in a neon matrix, acquired using the apparatus described above, is shown in Fig.~\ref{fig:spectrum}. We observe vibrational modes of linear carbon chains ranging in size from $\text{C}_3$ up to $\text{C}_{12}$, where peak assignments are based on previous matrix isolation spectroscopy results.~\cite{Weltner1964a,Weltner1964b,Thompson1971,Shen1989,Szczepanski1991,Smith1994,Kranze1995,Szczepanski1996,Kranze1996,Freivogel1997,Lapinski1999,Ding2000} As a consistency check, we have calculated correlation coefficients for all clusters with multiple absorption lines ($\text{C}_5$ to $\text{C}_{11}$) using data over a wide range of system parameters, and find $r > 0.995$ in all cases. This indicates that our peak assignment based on previous work is valid. A summary of the data depicted in Fig.~\ref{fig:spectrum} is given in the first three columns of Table~\ref{tab:carbon_lines}.
\begin{table}
\caption{\label{tab:carbon_lines} Representative data for $Q_{\text{Ne}} = 40~\text{sccm}$, $E_p = 2~\text{mJ}$, and $f_r = 7.5~\text{Hz}$. The center frequency of each absorption line, $\nu_0$, is given along with theoretical infrared intensities, $S$, peak optical depth per ablation pulse, $\Delta \tau_{pp}$, and the number of each carbon cluster produced per ablation pulse, $N_{pp}$. Quoted uncertainties are statistical.}
\begin{ruledtabular}
{\renewcommand{\arraystretch}{1}
\begin{tabular}{cccccc}
Molecule & Mode & $\nu_0$\footnote{Line centers are determined from a Gaussian fit to each peak, and uncertainties represent $95\%$ confidence bounds on the fit.} & $S$\footnote{Infrared intensities from Ref.~[\onlinecite{Hutter1994}].} & $\Delta \tau_{pp}$ & $N_{pp}$ \\
 &  & $(\text{cm}^{-1})$ & $(\text{km/mol})$ & $\left(10^{-6}\right)$ & $\left(10^{11}\right)$ \\
\hline
$\text{C}_3$ & $\nu_3$ & 2036.59(1) & 612.0 & 21.6(4) & $10.8(2)$ \\
\\
$\text{C}_4$ & $\nu_3$ & 1547.07(1) & 321.0 & 6.44(8) & $6.12(8)$ \\
\\
$\text{C}_5$ & $\nu_3$ & 2166.36(3) & 1648.8 & 20.4(3) & $3.78(6)$ \\
 & $\nu_4$ & 1444.24(3) & 97.3 & 1.38(6) & $4.3(2)$ \\
 \\
$\text{C}_6$ & $\nu_4$ & 1958.65(1) & 800.8 & 8.7(1) & $3.33(4)$ \\
 & $\nu_5$ & 1199.29(2) & 60.3 & $0.91(7)$ & $4.6(4)$ \\
 \\
$\text{C}_7$ & $\nu_4$ & 2135.08(2) & 2695 & 16.4(2) & $1.86(2)$ \\
 & $\nu_5$ & 1897.31(1) & 677.4 & 9.16(6) & $4.13(3)$ \\
 \\
$\text{C}_8$ & $\nu_5$ & 2067.69(1) & 1366.3 & 4.20(4) & $0.937(9)$ \\
 & $\nu_6$ & 1707.47(5) & 984.8 & 1.61(8) & $0.50(3)$ \\
 \\
$\text{C}_9$ & $\nu_5$ & 2081.13(3) & 2324 & 1.99(4) & $0.261(5)$ \\
 & $\nu_6$ & 2010.02(1) & 3030 & 22.1(2) & $2.22(2)$ \\
 & $\nu_7$ & 1602.74(1) & 376.5 & 1.68(5) & $1.36(4)$ \\
 \\
$\text{C}_{10}$ & -- & 2075.01(3) & -- & 1.97(4) & -- \\
 & -- & 1915.85(5) & -- & 1.24(9) & -- \\
 \\
 $\text{C}_{11}$ & $\nu_7$ & 1938.72(3) & -- & 2.88(6) & -- \\
 & $\nu_8$ & 1853.86(5) & -- & 1.75(5) & -- \\
 \\
$\text{C}_{12}$\footnote{Tentative assignment based on Ref.~[\onlinecite{Freivogel1997}].} & $\nu_8$ & 2003.47(8) & -- & 4.08(6) & -- \\
%$\text{C}_{11}$ & $\nu_7$ & 1938.5 & 13530~\cite{Botschwina2006} & 2.88(6) & 0.065(2) \\
% & $\nu_8$ & 1853.9 & 3378~\cite{Botschwina2006} & 1.75(5) & 0.159(5) \\
% \\
%$\text{C}_{12}$ & $\nu_8$ & 2003.1\footnote{Tentative} & 4794~\cite{Ding2000} & 4.08(6) & 0.260(4) \\
\end{tabular}}
\end{ruledtabular}
\end{table}
Note that the spectrum in Fig.~\ref{fig:spectrum} is acquired after 30 minutes of deposition onto the matrix window held at a temperature of about 5 K. Because 5 K is below the neon matrix annealing temperature of about 10 K, we conclude that the larger clusters observed spectroscopically are formed within the buffer-gas cell. This is in contrast to other matrix isolation studies of carbon clusters,~\cite{Weltner1964a,Weltner1964b,Thompson1971,Szczepanski1991,Szczepanski1996} where an additional annealing step was necessary to promote diffusive aggregation of smaller clusters into larger clusters.~\cite{Szczepanski1992,Grutter1997}

To quantify the number of each cluster present in a given spectrum, we start from Beer's law:
\begin{equation}\label{eq:BLlaw}
\tau(\nu) = \ln{\left[\frac{I_0(\nu)}{I(\nu)}\right]},
\end{equation}
where $I_0(\nu)$ and $I(\nu)$ are the incident and transmitted intensity, respectively, and $\tau(\nu)$ is the frequency-dependent optical depth of the sample (absorbance is related to optical depth by $A \approx 0.434\tau)$. From Beer's law, we define the integrated molar absorption coefficient, or infrared intensity $S$ as~\cite{Bernath2005}
\begin{equation}\label{eq:IRint}
S = \frac{1}{nl}\int d\nu~\tau(\nu),
\end{equation}
where $n$ is the molar concentration of the absorbing species, $l$ is the absorption length, and the integration is over the entire absorption line. Assuming that we probe an area of uniform density, we estimate the number of molecules in the matrix as
\begin{equation}\label{eq:N}
N \approx \left(\frac{N_A A_p}{S}\right)\Delta\nu~\tau_p,
\end{equation}
where $N_A$ is Avogadro's constant, $A_p$ is the approximate area of the buffer-gas beam at the matrix window, and the integral has been approximated as the product between the width of the absorption line, $\Delta\nu$, and the peak optical depth, $\tau_p$. For the calculations here, we use $A_p\sim5~\text{cm}^2$, and assume $\Delta\nu\sim1~\text{cm}^{-1}$ as a typical value for matrix isolated species.~\cite{Jacox2002}

Given Eq.~\eqref{eq:N}, we calculate the number of molecules produced per ablation pulse in the following way. We first extract $\tau_p$ for each observed peak in the spectrum after every two-minute deposition cycle, then use linear regression to calculate the time rate of change of $\tau_p$ for each peak. Finally, we convert this time rate of change to a change in peak optical depth per ablation pulse, $\Delta\tau_{pp}$, using the repetition rate of the laser. Thus, we arrive at the following expression for the number of molecules produced per ablation pulse:
\begin{equation}\label{eq:Npp}
N_{pp} \approx 3\times10^{19}\left(\frac{\Delta\tau_{pp}}{S}\right),
\end{equation}
where $S$ is in units of km/mol, and we have substituted $A_p$ and $\Delta\nu$ for the values given above. Table~\ref{tab:carbon_lines} shows the results of this analysis for a representative data set with $Q_{\text{Ne}} = 40~\text{sccm}$, $E_p = 2~\text{mJ}$, and $f_r = 7.5~\text{Hz}$. Note that this calculation depends on knowledge of the infrared intensity for a given molecular transition. As this parameter is difficult to measure directly, we rely on results from density functional theory calculations for linear $\text{C}_3$ to $\text{C}_9$.~\cite{Hutter1994} We neglect $\text{C}_{10}$ to $\text{C}_{12}$ in our analysis as we have been unable to find consistent theoretical values of $S$ for the observed absorption lines. For $\text{C}_5$ to $\text{C}_8$, we find reasonable agreement (i.e., same order of magnitude) for the value of $N_{pp}$ calculated from different absorption lines of the same molecule, which implies good relative accuracy of the theoretical infrared intensities. However, for $\text{C}_9$, the value of $N_{pp}$ calculated for the $\nu_5$ mode is off by an order of magnitude from the corresponding values for the $\nu_6$ and $\nu_7$ modes. This inconsistency likely stems from a neglect of electrical anharmonicity in the theoretical method used to determine $S$.

\begin{figure}
\centering
\subfloat{\includegraphics[scale=1]{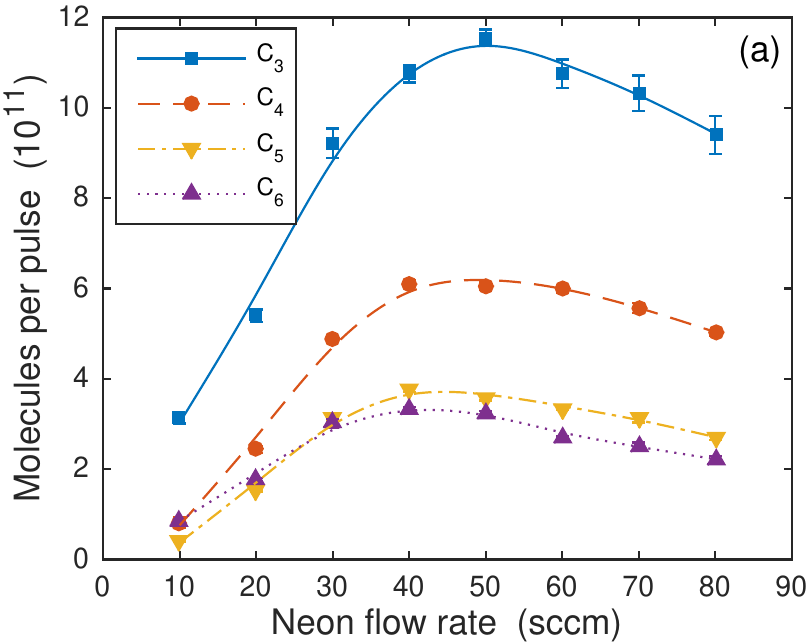}}\\
\subfloat{\includegraphics[scale=1]{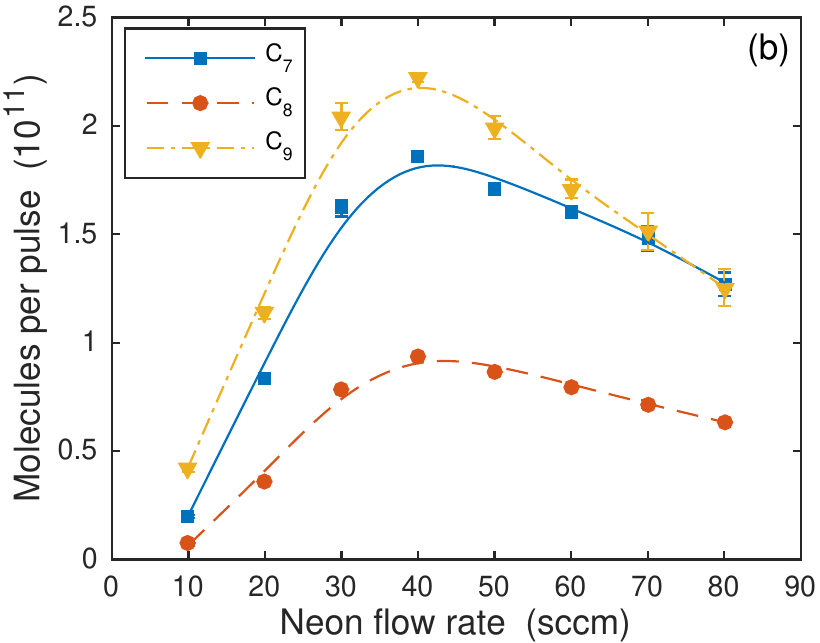}}
\caption{\label{fig:mols_flow} Number of each carbon cluster produced per pulse for (a) $\text{C}_3$ to $\text{C}_6$ and (b) $\text{C}_7$ to $\text{C}_9$ as a function of neon flow rate for $E_p = 2~\text{mJ}$ and $f_r = 7.5~\text{Hz}$. Error bars represent statistical uncertainty propagated from the determination of $\Delta\tau_{pp}$. Lines are guides for the eye.}
\end{figure}

We now apply the analysis described above to investigate carbon cluster production in the buffer-gas cell as a function of the system parameters. For clusters that have multiple absorption lines, we use the line with the largest infrared intensity (corresponding to largest peak optical depth) to calculate $N_{pp}$. In Fig.~\ref{fig:mols_flow}, the number of each carbon cluster produced per pulse is plotted versus the flow rate of neon buffer gas.
For all detected clusters, we observe a linear increase in production with flow rate up to about $Q_{\text{Ne}} = 40~\text{sccm}$. This increase then saturates and begins to decrease for higher flow rates, an effect that has been observed in a number of other buffer gas experiments.~\cite{Sushkov2008,Lu2008,Skoff2011,Hutzler2011,Bulleid2013}

The effect of neon flow rate on the number of each cluster detected in the matrix can be understood by considering the two relevant timescales that characterize the buffer-gas cell dynamics: the molecule diffusion time to the cell walls, $t_d$, and the cell pumpout time, $t_p$. If $t_d \ll t_p$ then molecules produced by ablation diffuse to the cell walls and are lost before they are extracted from the cell. On the other hand, if $t_d \gg t_p$, we expect that the molecules are extracted from the cell before being lost to the cell walls resulting in a larger number of molecules detected inside the matrix. The cell pumpout time is governed by the conductance of the cell aperture, and is therefore purely geometrical. The diffusion time, however, depends on the density of neon buffer gas inside the cell, which is linearly proportional to the flow rate. Thus, we characterize the extraction behavior using the dimensionless quantity~\cite{Hutzler2012} $\gamma_{\text{cell}}\equiv t_d/t_p$. It is typically assumed that $t_d$ is dominated by the lowest order diffusion mode in the cell (a good approximation at low buffer-gas density~\cite{Skoff2011}), and increases linearly with the neon density in the cell.~\cite{Hasted1972} For our system parameters, we find $\gamma_{\text{cell}} \approx 1$ for $Q_{\text{Ne}} \approx 25~\text{sccm}$, which corresponds to $t_d = t_p \approx 8~\text{ms}$. This correlates well with the observed linear increase in molecules per pulse up to about $Q_{\text{Ne}} = 30-40~\text{sccm}$ shown in Fig.~\ref{fig:mols_flow}. As $\gamma_{\text{cell}}$ becomes much larger than unity, we would expect the number of detected clusters to saturate as all molecules should be extracted from the cell before diffusing to the cell walls. However, at high buffer-gas densities (i.e., high flow rate) the diffusion process is no longer dominated by the lowest order mode.~\cite{Skoff2011} Since higher-order diffusion modes have smaller time constants,~\cite{Hasted1972} the result is an effective decrease in $\gamma_{\text{cell}}$, and therefore in the extraction efficiency. Additionally, it is possible that the neon flow becomes turbulent,~\cite{Bulleid2013} and negatively impacts the extraction efficiency in a similar way. We observe these effects as a reduction in the number of each carbon cluster detected in the matrix at neon flow rates in excess of about $40-50~\text{sccm}$.

\begin{figure}
\centering
\subfloat{\includegraphics[scale=1]{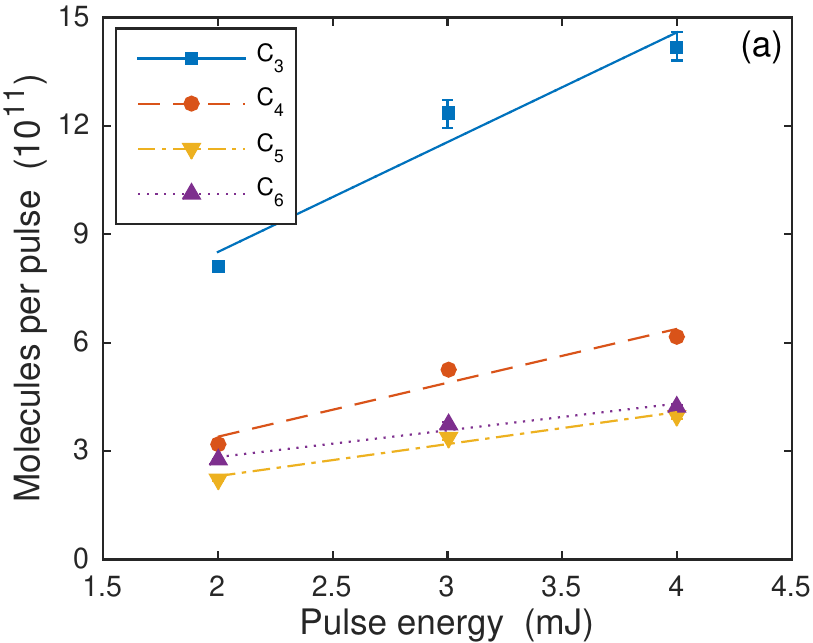}}\\
\subfloat{\includegraphics[scale=1]{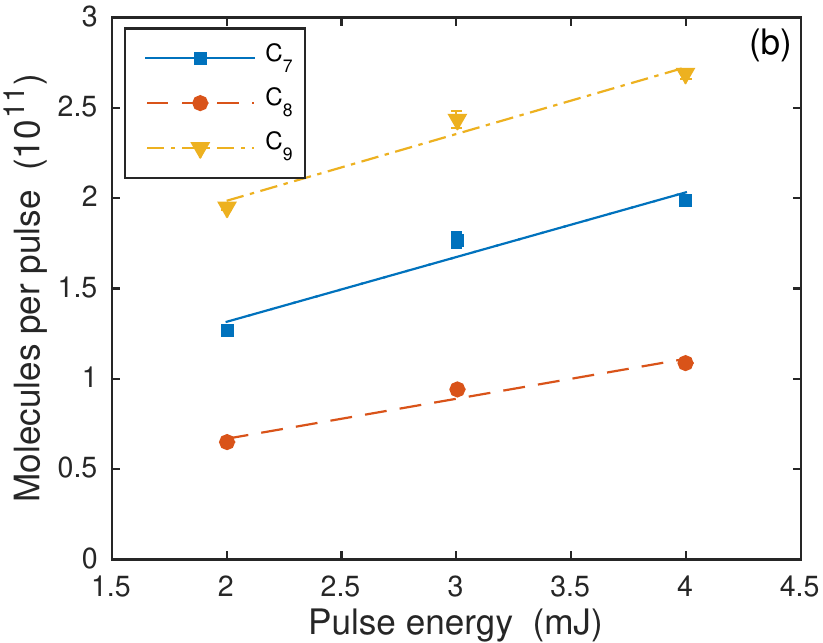}}
\caption{\label{fig:mols_enrg} Number of each carbon cluster produced per pulse for (a) $\text{C}_3$ to $\text{C}_6$ and (b) $\text{C}_7$ to $\text{C}_9$ as a function of ablation pulse energy for $Q_{\text{Ne}} = 40~\text{sccm}$ and $f_r = 7.5~\text{Hz}$. Error bars represent statistical uncertainty propagated from the determination of $\Delta\tau_{pp}$. Lines are guides for the eye.}
\end{figure}

In addition to flow rate, carbon cluster production as a function of the ablation laser pulse energy and repetition rate is investigated, the former of which is plotted in Fig.~\ref{fig:mols_enrg}.
We observe a linear increase in the production of all carbon clusters with increasing ablation pulse energy up to $E_p = 4~\text{mJ}$, at which point a substantial amount of carbon dust was present in the vacuum chamber. Therefore, we did not explore higher pulse energies. We observed no variation in cluster production rate for laser repetition rates of $f_r = 3,~5,~\text{and}~7.5~\text{Hz}$, implying local heating of the substrate does not effect cluster production at these low repetition rates.

\section{\label{sec:conclusion}Conclusion}
We have produced carbon clusters $\text{C}_3$ to $\text{C}_{12}$ using a cryogenic buffer-gas beam source. For each of the $\text{C}_3$ to $\text{C}_9$ molecules, we estimate a production rate on the order of $10^{11}-10^{12}$ molecules per pulse of the ablation laser based on infrared intensities determined from density functional theory calculations. Similar to other buffer-gas beam experiments, we observe an initial linear increase in molecule production with neon flow rate followed by a decrease at high flow rates (i.e., high neon density in the cell). We attribute this decrease at high flow rates to a reduction in the diffusion time for molecules to be lost to the cell walls.

Based on the direct observation of large numbers of carbon clusters containing more than a few atoms, the buffer-gas beam source may be a viable method for the production of more exotic cluster systems such as metal oxides and carbides. Furthermore, given the increased number of collisions in comparison to a supersonic expansion source, we anticipate the buffer-gas beam to be a useful source for the production of vibrationally cold cluster systems. Indeed, the ability to cool complex molecules to their vibrational ground state will remove much of the ambiguity associated with the spectral assignment process, and could provide validation for the large variety of theoretical techniques currently being used in conjunction with experiments to determine molecular structures.

\acknowledgements{We would like to thank the Doyle/Patterson group at Harvard for providing the buffer-gas cell used in this work, and for useful discussions regarding its operation. This work was supported by the National Science Foundation Physics Frontier Center (Grant No. PHY1125844) and through the University of Colorado's Innovative Seed Grant Program.}

\end{document}